\documentclass[11pt,twoside]{article}      
\usepackage{asp2004}
\usepackage{psfig}
\usepackage{epsf}
\usepackage{graphics}
\begin{document}

\title{Mapping WR\,140 from infrared spectroscopy and imaging}

\author{P. M. Williams}
\affil{Institute of Astronomy, Royal Observatory, Edinburgh, EH9 3HJ, U.K.}
\author{W. P. Varricatt}
\affil{Joint Astronomy Centre, 660 N. A`oh\=ok\=u Place, Hilo, HI 96720, U.S.A.}
\author{A. P. Marston}
\affil{European Space Research and Technology Center (ESTEC), European Space 
Agency, Keplerlaan 1, AG NL-2200 Noordwijk, Netherlands}
\author{N. M. Ashok}
\affil{Physical Research Laboratory, Navrangpura, Ahmedabad, India 380009}

\begin{abstract}
Observations of the 1.083-$\mu$m He\,{\sc i} line in WR\,140 (HD 193793) show 
P-Cygni profiles which varied as the binary system went through periastron 
passage. A sub-peak appeared on the normally flat-topped emission component 
and then moved across the profile consistent with its formation in the 
wind-collision region. Variation of the absorption component provided 
constraints on the opening angle ($\theta$) of the wind-collision region.  
Infrared (2--10-$\mu$m) images observed with a variety of instruments in 2001--04 
resolve the dust cloud formed in 2001, and show it to be expanding at a constant 
rate. Owing to the high eccentricity of the binary orbit, the dust is spread 
around the orbital plane in a `splash' and we compare the dust images with 
the orientation of the orbit. 
\end{abstract}

\thispagestyle{plain}

\section{Infrared spectroscopy and the 1.083-\boldmath $\mu$m He\,{\sc i} P Cygni profile}


The WC7+O4-5 binary WR\,140 (HD 193793) is the archetypal episodic dust-forming 
Wolf-Rayet Colliding Wind Binary. It is particularly luminous in X-rays and was the 
first WR system to show non-thermal radio emission and episodic dust formation. 
Variations in its X-ray, radio and infrared properties were linked to its 
binary motion by Williams et al. (1990), and it has since been the subject of 
theoretical and observational studies of colliding-wind phenomena. As part of 
the campaign planned for the 2001 periastron passage, we observed a series of 
near-infrared spectra having resolutions $R \simeq 1000$ using the United Kingdom 
Infrared Telescope (UKIRT), Hawaii, and the Mt Abu Infrared Telescope, India 
(Varricatt, Williams \& Ashok 2004). The 1.083-$\mu$m He\,{\sc i} line was 
also observed at $R = 4700$ to look for colliding-wind effects (cf. Stevens 
\& Howarth 1999). Previous observations of WR\,140 (e.g. Eenens \& Williams 1994) 
had shown flat-topped profiles for this line, but they were taken far from 
periastron. 
Our observation at $\phi = 0.96$, however, showed the appearance of a strong, 
blue-shifted sub-peak on the emission profile (Fig.\,1), which we interpret 
as being formed in a shell of compressed WC7 stellar wind material flowing along 
the wind-collision region (WCR). The shape of the WCR can be approximated (e.g. 
Usov 1992., L\"uhrs 1996) by a hollow cone symmetric about the axis joining the 
stars and having its apex towards the WC7 star, which has the higher 
mass-loss rate.

\begin{figure}[!ht]
\plottwo{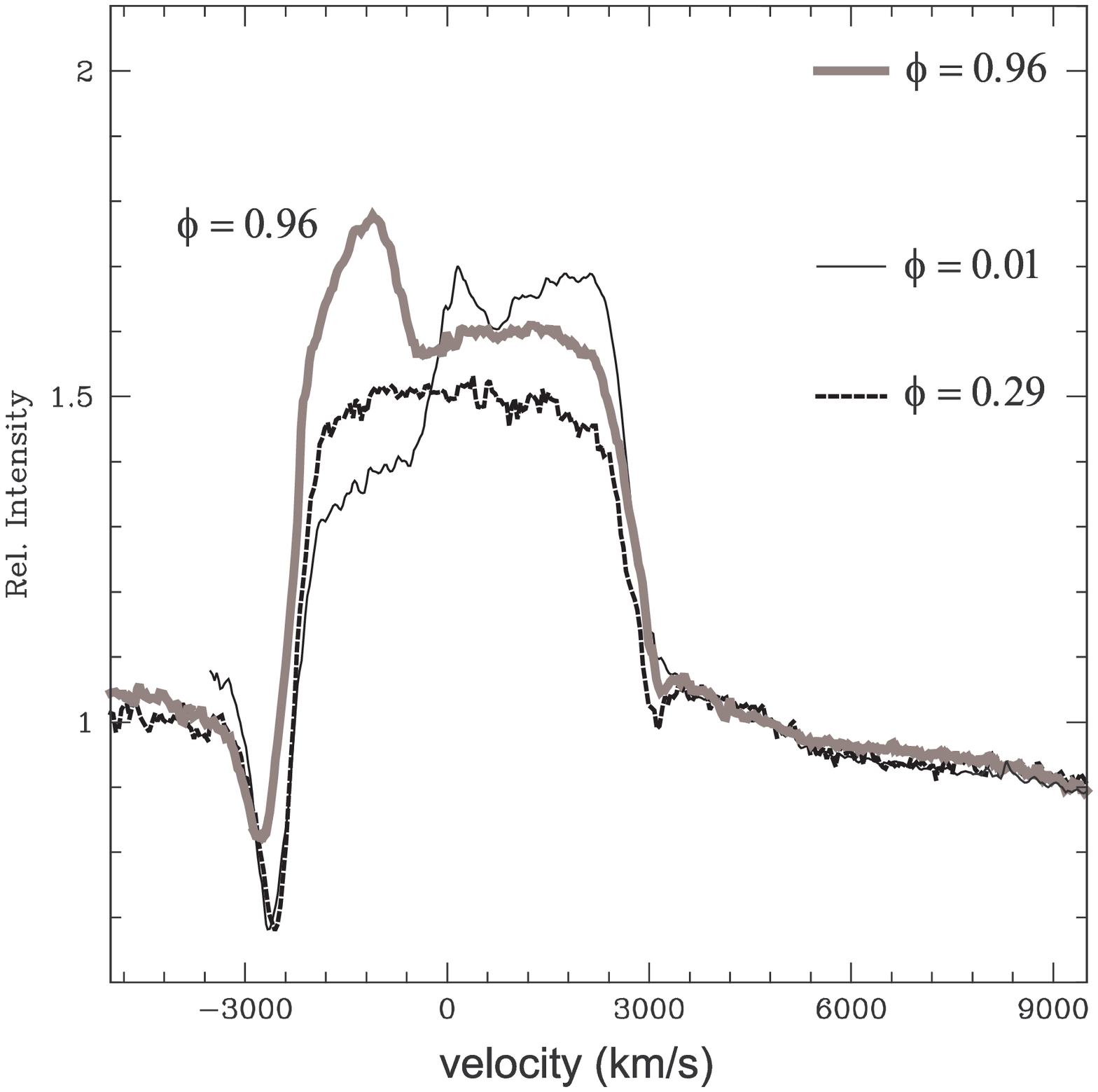}{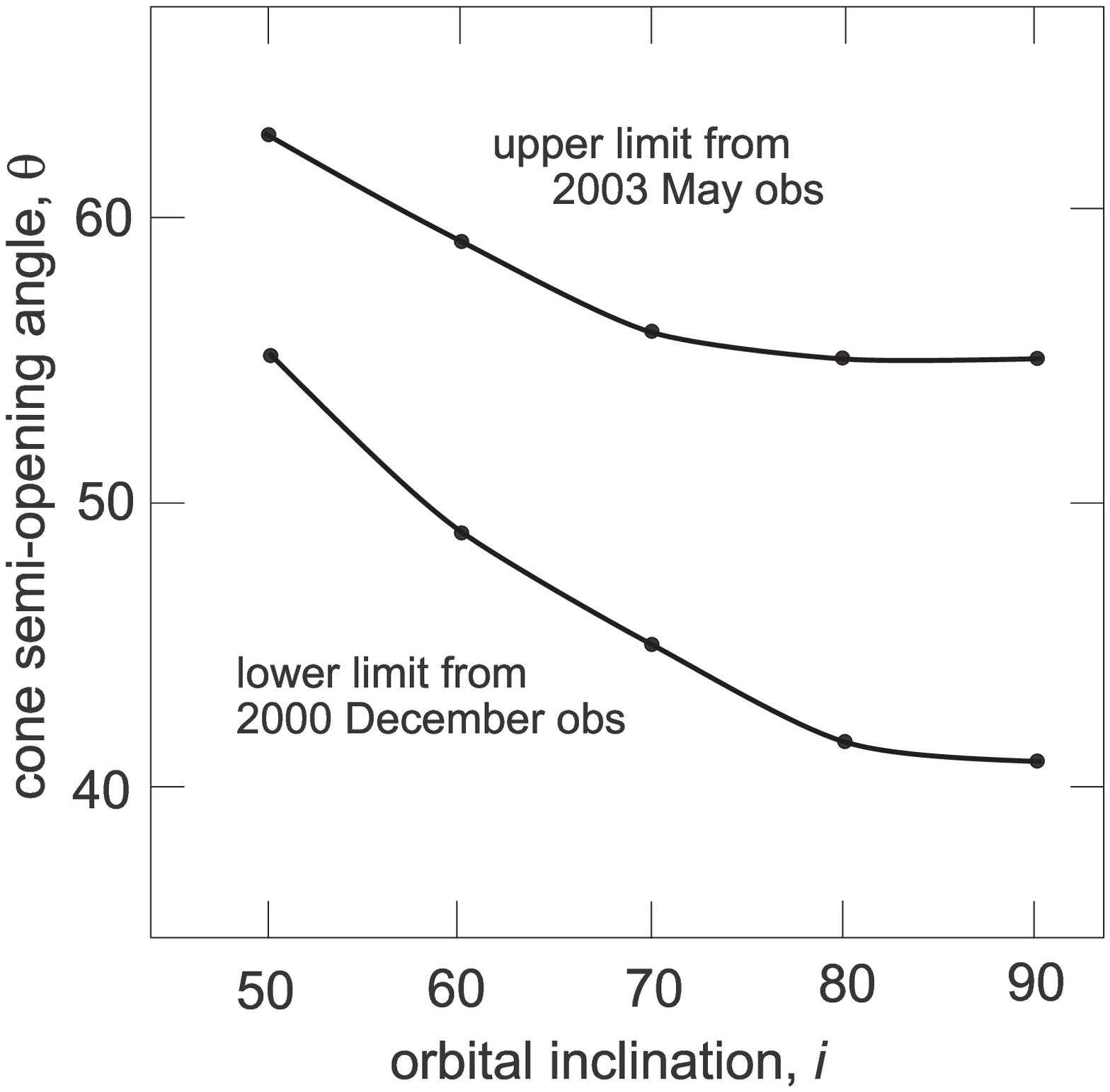}
\caption{Left: profiles of the 1.083-$\mu$m He\,{\sc i} line in 2000 October 
($\phi = 0.96$), 2001 March ($\phi = 0.01$) and 2003 ($\phi = 0.29$), which 
is typical of most of the orbit). 
Right: constraints on the opening angle ($\theta$) of the wind-collision 
region as a function of orbital inclination set by the varying strength 
of the absorption component of the He\,{\sc i} line.}
\end{figure}

At $\phi = 0.96$, the system is near conjunction with the O star in front of 
the WC7 star, so we expect the compressed WC7 material to be flowing towards 
us. At the same time, we are viewing the two stars mostly through the wind 
of the O star, accounting for our observation (Fig.\,1) that the absorption 
component of the He\,{\sc i} line is weaker than at phases when the stars 
are observed through the He-rich WC7 stellar wind. At this phase, the angle 
$\psi$ between our sightline and the axis joining the two stars is smaller 
than the opening angle ($\theta$) of the wind-collision `cone'.

The value of $\psi$ varies round the orbit and can be calculated for the 
phases of our observations for a range of values of the inclination 
so we can use successive measurements of the strength of the absorption 
component to explore permitted values of $\theta$ and inclination. 
At the same time, we model the variation of the radial velocity (RV) of the subpeak, including 
its width, following L\"uhrs (1997) but with the difference that, instead of 
treating the velocity of the compressed wind as a free parameter to be fit, 
we calculate it from those of the WC7 and O star winds and $\theta$ following 
Cant\'o, Raga \& Wilkins (1996). 
This models the movement of the sub-peak to the red end of the profile by 
the time of our first post-periastron observation ($\phi = 0.01$, Fig.\,1). 
At this time, the absorption component was strong because we observed both 
stars through the WC7 stellar wind. Two years later, by $\phi = 0.29$, the 
emission sub-peak had vanished as the stars had moved further apart, but 
the absorption component was still strong. This puts a useful upper limit 
on $\theta$ (Fig.\,1). Modelling the RVs of the sub-peak indicates an 
orbital inclination $i \simeq 65\deg\pm10\deg$, implying (Fig.\,1) 
$\theta \simeq 53\deg\pm10\deg$.

\section{Infrared imaging of the dust emission}

We imaged WR\,140 in the infrared with four different instruments: 
PHARO +AO on the Hale telescope, INGRID+AO (NAOMI) on the William Herschel 
Telescope (WHT), and UIST and Michelle, both on UKIRT. The UKIRT observations 
were made at longer wavelengths, where the contrast between dust and stellar 
emission was greater, especially as the dust cooled, but the resolution is 
lower. Single stars were observed to determine the image psfs and the images 
of WR\,140 were restored using maximum entropy methods. 
A test of the validity of our procedures is provided by two images of 
WR\,140 (Fig.\,2) observed at about the same time with two different 
instruments and reduced with different software packages. The basic 
structures are similar, with a `bar' of dust emission to the south 
and another feature  to the east. Similar structures are evident in all the 
images, e.g. a 3.99-$\mu$m image observed with UIST in 2003 (Fig.\,3). 

\begin{figure}[!ht]
\plottwo{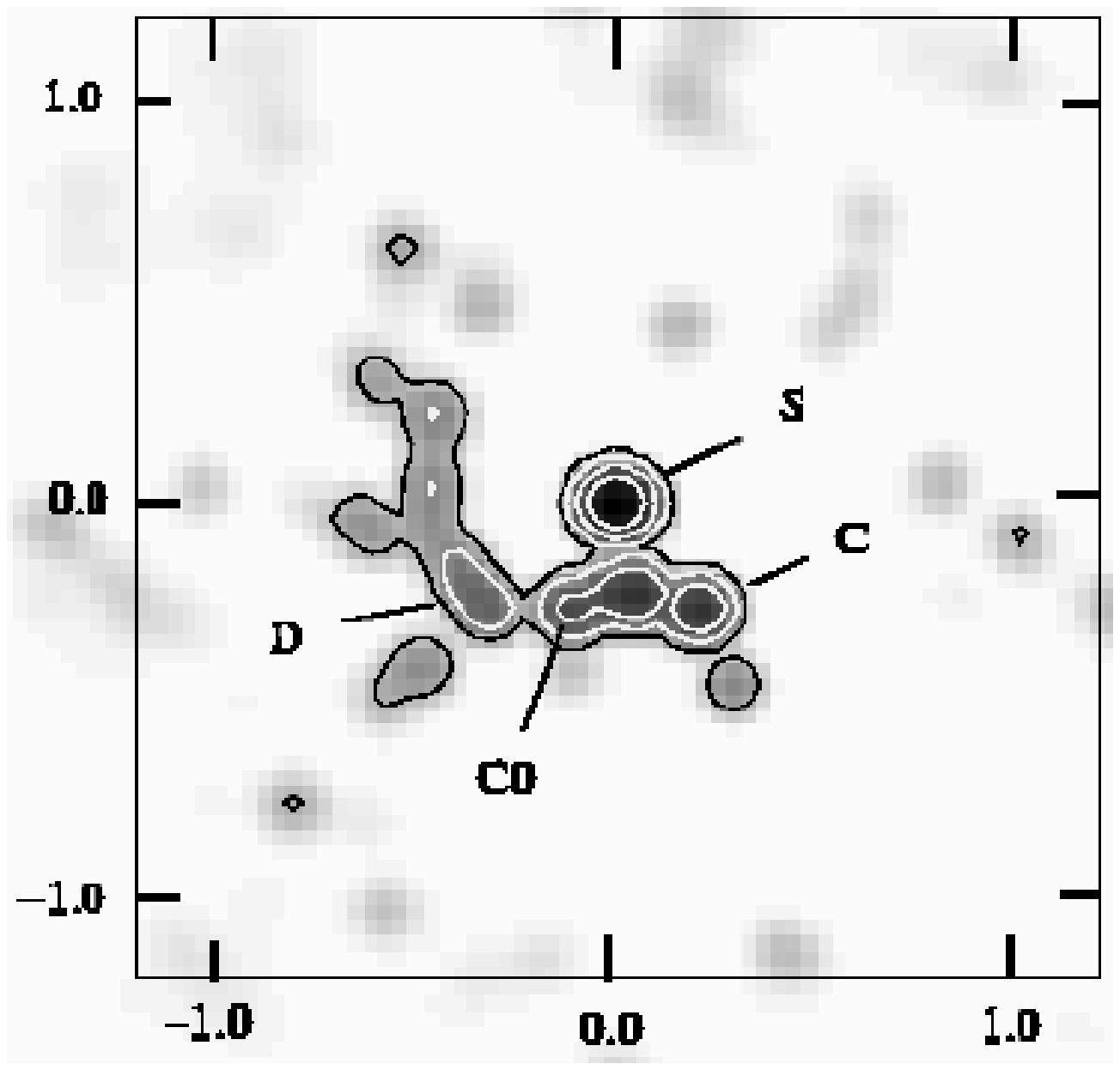}{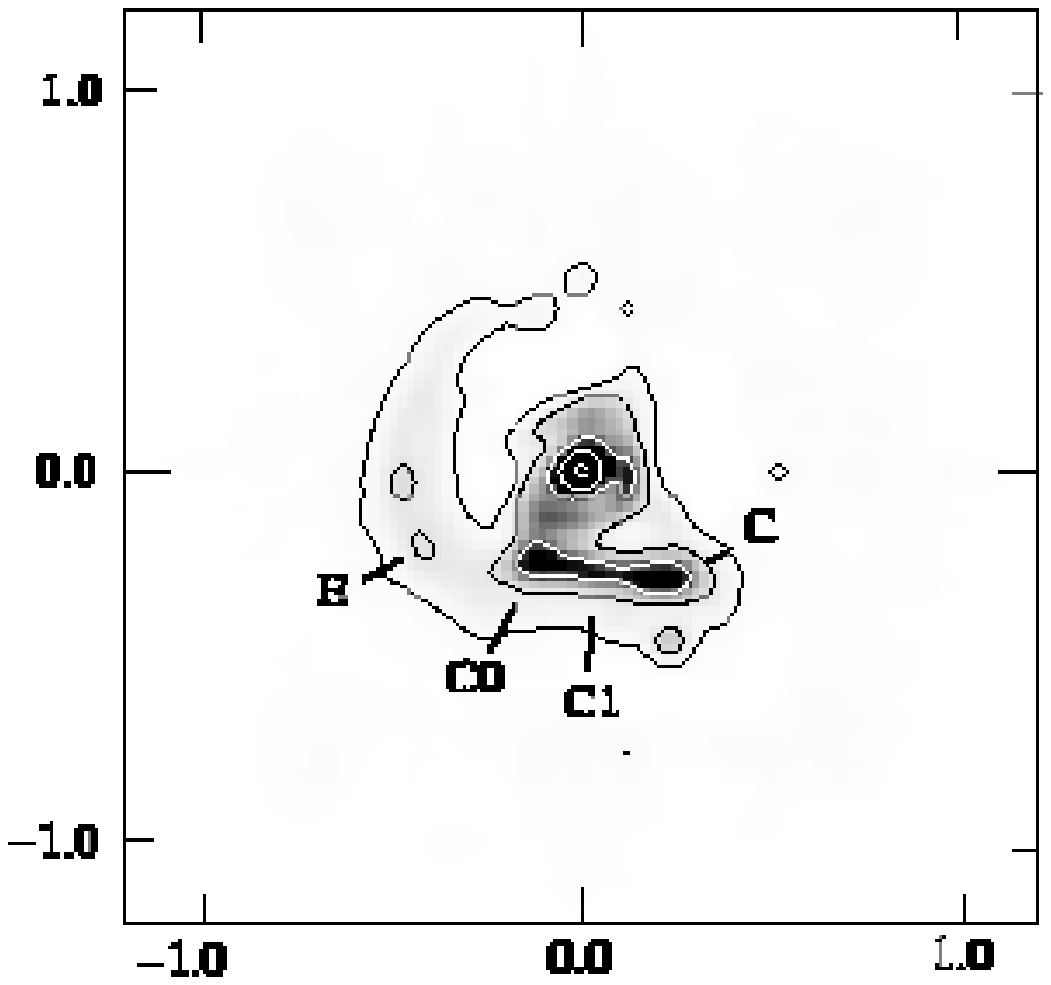}
\caption{Two 2-$\mu$m images of WR\,140 observed in 2002 July with different 
instruments: NAOMI/INGRID on the WHT (left) and AO/PHARO system on the Hale (right). 
Both images have NE at top left and the scale is arcsec from the central star (`S'). 
Dust emission knots (`C', `D' and `E') correspond to those identified in the 
2001 images of Monnier et al. (2002).}
\end{figure}

We can use the contemporaneous 3.6-$\mu$m and 3.99-$\mu$m UIST observations to 
measure the infrared colours of the emission features relative to that of the star, 
confirming that the the features are 0.4 mag. redder in [3.6]--[3.99] than the star, 
consistent with their being heated dust.
We identified several knots of dust emission with those identified in the  
aperture-masking images observed in 2001 by Monnier, Tuthill \& Danchi (2002) 
from the similarity of their position angles (P.A.) relative to the star, 
and use these and other knots to track the expansion of the dust cloud. 
The observations are summarized in Table 1, together with the P.A.s and  
radial distances ($r$) of the emission knot, `C'. These distances, together 
with those from the images of Monnier et al. and measured from the image 
given by Tuthill et al. (2003), and similar data for knot `E' to the east, 
are plotted against date in Fig.\,3, and show remarkably linear expansion. 

\begin{table} 
\caption{Log of imaging observations of WR\,140 with position angle and 
distance of dust-emission knot `C' from the star}
\smallskip
\begin{center}
{\small
\begin{tabular}{lcccccc}
\tableline
\noalign{\smallskip}

Instrument      & Date    & Phase & Scale  &$\lambda$obs & P.A.    & r   \\
                &         &       & (mas) & ($\mu$m)     & ($\deg$) & (mas)\\
\noalign{\smallskip}
\tableline
\noalign{\smallskip}

Hale: PHARO+AO  & 2001.68 & 0.073 & 25   & 2.2      &       &       \\
Hale: PHARO+AO  & 2002.31 & 0.153 & 25   & 2.2      &  216  & 278   \\
WHT: INGRID+AO  & 2002.51 & 0.178 & 26.3 & 2.27     &  219  & 341   \\
Hale: PHARO+AO  & 2002.56 & 0.184 & 25   & 2.2      &  217  & 352   \\
UKIRT: UIST     & 2002.89 & 0.226 & 16.5 & 3.6      &  217  & 498   \\
UKIRT: UIST     & 2002.89 & 0.226 & 16.5 & 3.99     &  216  & 451   \\
UKIRT: UIST     & 2003.42 & 0.293 & 16.5 & 3.6      &  210  & 630   \\
UKIRT: UIST     & 2003.42 & 0.293 & 16.5 & 3.99     &  217  & 601   \\
UKIRT: Michelle & 2004.25 & 0.397 & 210  & 10.52    &  215  & 874   \\
UKIRT: UIST     & 2004.49 & 0.426 & 16.5 & 3.99     &  214  & 948   \\
UKIRT: UIST     & 2004.49 & 0.426 & 16.5 & 4.68     &  218  & 928   \\
\noalign{\smallskip}
\tableline
\end{tabular}
}
\end{center}  
\end{table}

\begin{figure}[!ht]
\plottwo{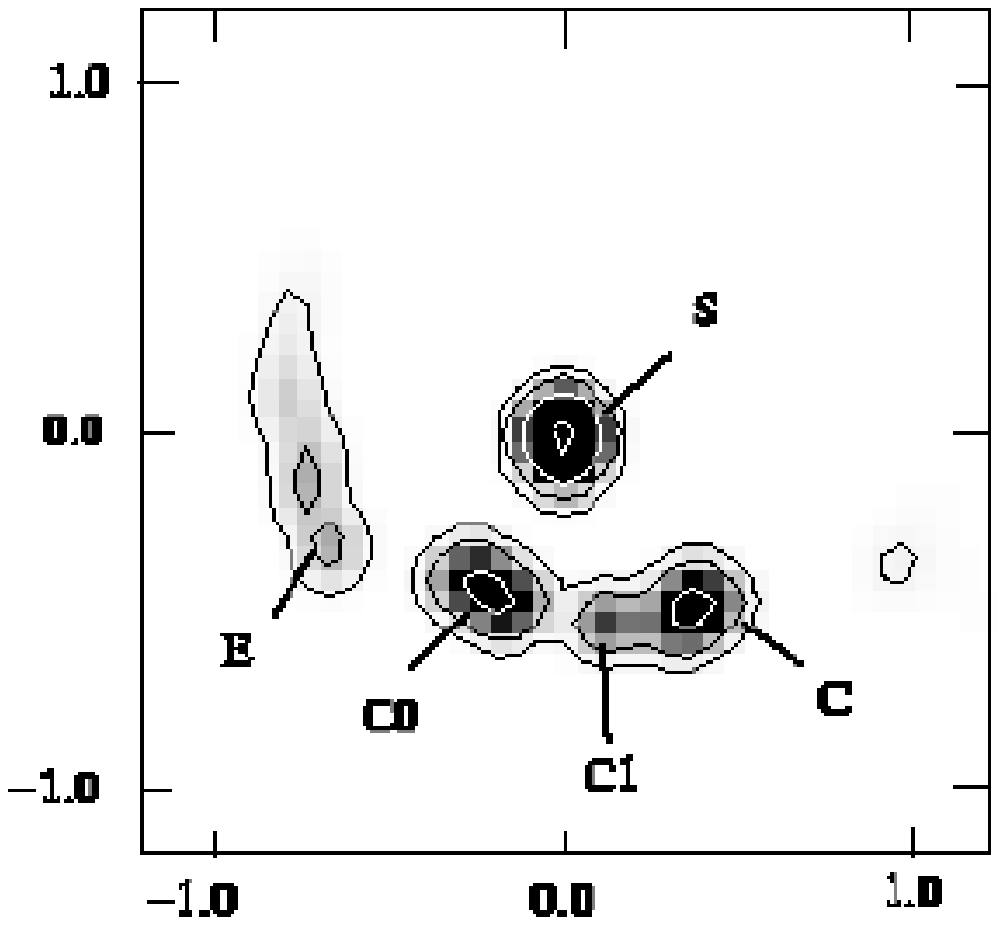}{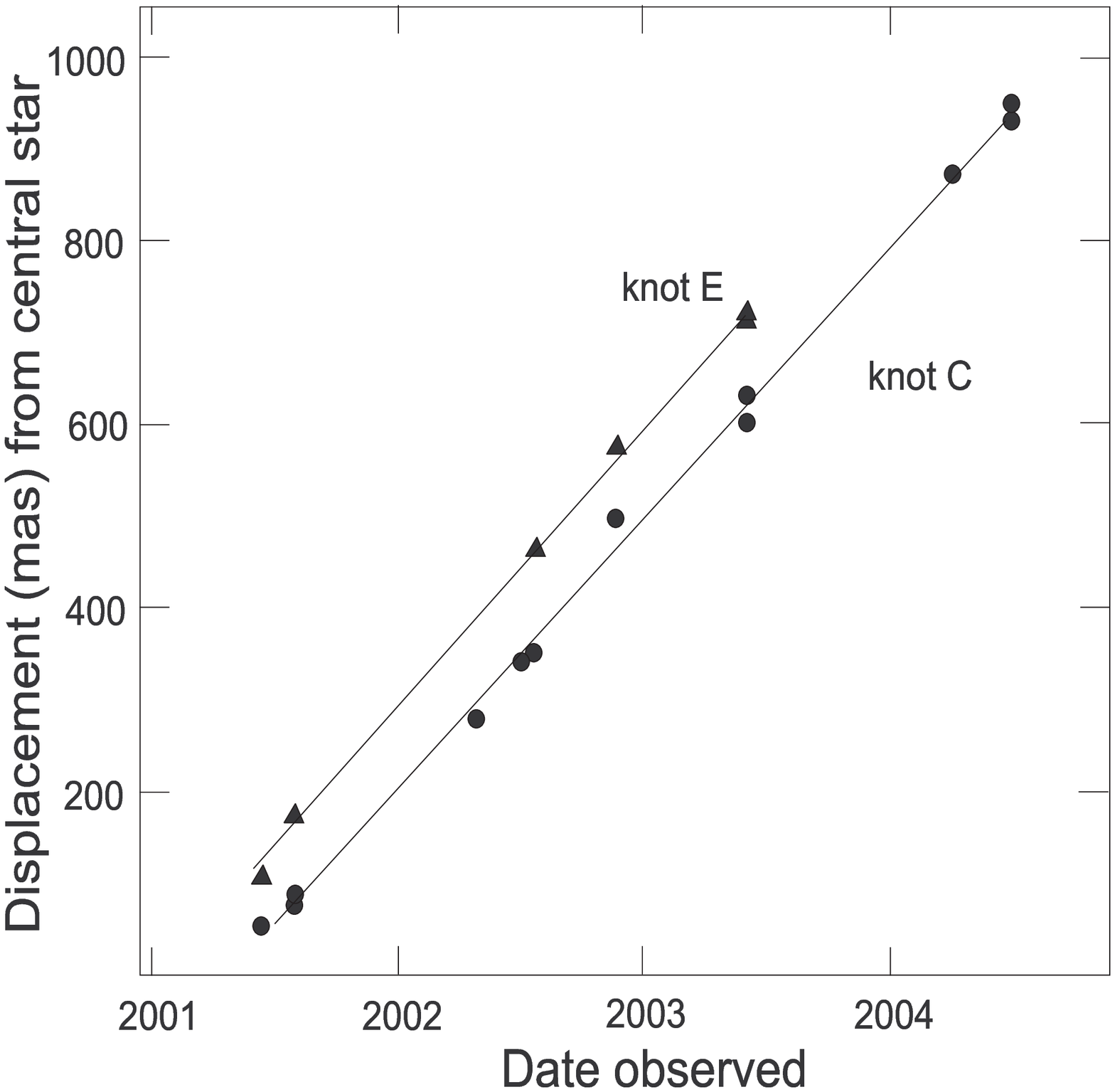}
\caption{Left: A 3.99-$\mu$m image of WR\,140 observed on 2003 June 4 with UIST  
on UKIRT showing homologous expansion of the dust emission features observed in 
earlier images. The resolution is lower than in Fig.\,2 owing to the longer 
wavelength. Right: proper motions of knots `C' and `E' of dust emission measured 
from our images and the 2001 images of Monnier et al. and Tuthill et al. (2003).}
\end{figure}

\begin{table}[!ht]
\caption{Proper motions, projected velocities and ejection dates of dust-emission 
 `knots'.}
\smallskip
\begin{center}
{\small
\begin{tabular}{lllll}
\noalign{\smallskip}
\tableline
\noalign{\smallskip}
Knot  & P. M.       & Proj. vel. &  Date started  &  Phase started \\
      & (mas/y)     & (km/s)     &                &          \\
\noalign{\smallskip}
\tableline
\noalign{\smallskip}
C     & 294$\pm$7   & 2575$\pm$63  & 2001.28$\pm$0.07 & 0.023$\pm$0.009 \\
C0    & 240$\pm$17  & 2108$\pm$151 & 2001.34$\pm$0.21 & 0.030$\pm$0.027 \\
D     & 313$\pm$18  & 2745$\pm$158 & 2001.22$\pm$0.11 & 0.015$\pm$0.014 \\
E     & 304$\pm$2   & 2664$\pm$14  & 2001.03$\pm$0.01 & 0.991$\pm$0.002 \\
\noalign{\smallskip}
\tableline
\end{tabular}
}
\end{center}
\end{table}  

The proper motions of selected knots from linear fits to the distances of 
selected knots are given in Table 2, together with projected velocities 
adopting the revised distance of 1.85 kpc (Dougherty et al., this meeting). 
These velocities are comparable to that (2470 km/s) of the compressed WC7 
wind material responsible for the He\,{\sc i} sub-peaks and presumably the 
gas in which the dust formed. The observation of relatively high projected 
velocities for three of the four well observed knots suggests either that 
they are clumps moving in the plane of the sky or, more probably, that they 
are limb-brightened edges of a hollow dust cloud and not physical clumps. 

To model the dust-emission images, we assume that the dust moves radially 
along the projection of the WCR and therefore need to know the changing 
configuration of the WCR during dust formation. From the NIR light curves 
(Williams 1990), we infer that dust formation occurs for only $2-3\%$ of 
the period on either side of periastron. The high eccentricity of the 
orbit (Marchenko et al. 2003), however, means that the axis of the WCR 
moves through more than $180\deg$ in this short time and, allowing for 
$\theta \simeq 53\deg$, that the dust is spread around three-quarters of 
the obital plane in a small fraction of the period. This will not give a 
dust `spiral', nor a `jet', but a `splash'. The orientation of this 
`splash' on the sky using the orientation of the orbit from values of 
$\Omega$ and $i$ derived by Dougherty et al. has the dust starting to 
the east and running clockwise round the star to finish to the south, 
making the southern `bar' the most recently formed dust. The absence of 
significant dust emission to the NW, the projected axis at the time 
of periastron passage, suggests that dust formation is quenched at the 
very closest separation.

\acknowledgments{We gratefully acknowledge the Service Observing Programmes 
of the Isaac Newton Group (WHT) and the Joint Astronomy Centre (UKIRT).}

\end{document}